\documentclass[a4paper,10pt,twocolumn]{article}
\usepackage{inputenc}
\usepackage{tgtermes}
\usepackage{lipsum}
\usepackage{abstract}
\usepackage[compact]{titlesec}
\usepackage{titling}
\usepackage{dblfloatfix}
\usepackage{mathtools}
\usepackage[usenames, dvipsnames]{color}
\usepackage{soul}
\usepackage{amsmath,amssymb}
\usepackage{mathptmx}
\usepackage[symbol,hang,flushmargin]{footmisc}

\usepackage{caption}
\usepackage{tabu}
\usepackage{float}
\usepackage[export]{adjustbox}[2011/08/13]
\usepackage{multirow}
\usepackage{array}

\usepackage{upgreek}
\usepackage{booktabs}
\usepackage{fancyhdr}
\usepackage{nicefrac}
\usepackage{graphicx}
\usepackage[super, comma, sort&compress]{natbib}
\usepackage{authblk}
\usepackage{xspace}
\usepackage{pifont}
\usepackage[a4paper, total={6.8in, 10in}]{geometry}

\definecolor{myblue}{rgb}{0.03, 0.23, 0.63}
\newcommand{\ch}[1]{{\color{black}#1}}

\titleformat{\section}
{\normalfont\Large}{\; \; \thesection.}{1em}{}

\titleformat{\subsection}
{\normalfont\large}{\thesubsection.}{1em}{}

\newcolumntype{P}[1]{>{\centering\arraybackslash}p{#1}}

\newcommand*{\citen}[1]{%
  \begingroup
    \romannumeral-`\x 
    \setcitestyle{numbers}%
    \cite{#1}%
  \endgroup   
}

\makeatletter
\def\@maketitle{%
  \newpage
  \vspace*{-\topskip}     
  \begingroup\centering   
  \let \footnote \thanks
  \hrule height \z@       
    {\LARGE \@title \par}%
    \vskip 1.5em 
    {\large
      \lineskip .5em 
      \begin{tabular}[t]{c}%
        \@author
      \end{tabular}\par}%
    \vskip 1em 
    {\large \@date}%
  \par\endgroup           
  \vskip 1.5em            
}
\makeatother

\titlelabel{\quad \thetitle. \quad}

\title{Enhancement of a silicon waveguide single photon source 
by temporal multiplexing}

\author[]{Jeremy C. Adcock\thanks{ \noindent  jerad@fotonik.dtu.dk}}
\author[]{ Davide Bacco}
\author[]{Yunhong Ding\thanks{ \noindent  yudin@fotonik.dtu.dk}}
\affil[]{\small{Center for Silicon Photonics for Optical Communication (SPOC), 

Department of Photonics Engineering, Technical University of Denmark, Lyngby, Denmark}}

\date{October 2021}

\begin{document}

\pagestyle{fancy}
\lhead{J.C. Adcock et al.}
\chead{Enhancement of a silicon waveguide single photon source by temporal multiplexing}
\rhead{\thepage}
\setlength{\headwidth}{\textwidth}
\fancyfoot{}

\renewcommand{\abstractname}{}   
\renewcommand{\figurename}{Fig.}

\maketitle
\vspace{-1.5cm}{

\section*{\textbf{\large{ Abstract}}}
Efficient generation of single photons is one of the key challenges of building photonic quantum technology, such as quantum computers and long-distance quantum networks. 
Photon source multiplexing---where successful pair generation is heralded by the detection of one of the photons, and its partner is routed to a single mode output---has long been known to offer a concrete solution, with output probability tending toward unity as loss is reduced. 
Here, we present a temporally multiplexed integrated single photon source based on a silicon waveguide and a low-loss fibre switch and loop architecture, which achieves enhancement of the single photon output probability of $4.5 \pm 0.5$, while retaining $g^{(2)}(0) = 0.01$.
}

\begin{figure*}[h]
    \includegraphics[width=1\textwidth]{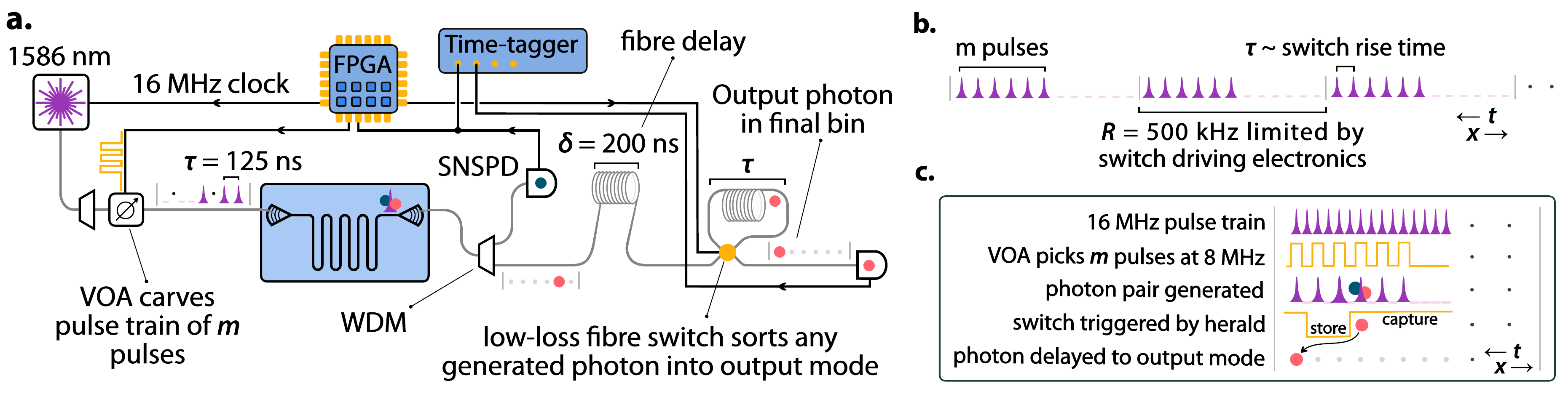}
    \centering
    \caption{Experimental set up and temporal-loop multiplexing dynamics. \textbf{a.} Schematic of our experiment. Photon pairs are probabilistically generated in each time bin via SFWM in a silicon waveguide. Detection of one of the pair causes the other to be switched into a fibre loop, routing the signal photon to the output time bin and boosting the output probability. \textbf{b.} A pulse train defines $m$ single temporally independent single photon sources. \textbf{c.} Pulse picking scheme and switching logic for temporal-loop multiplexing}
    \label{fig:fibre_plex_setup}
\end{figure*}


\section{Introduction}
\thispagestyle{empty}

Photonics is poised to play a key role in future quantum technologies, both in quantum computers and the quantum networks that will connect them. 
In the last decade, progress in photonic quantum information processing has accelerated dramatically, with historic challenges such as achieving near-unity photon purity recently being overcome\cite{graffitti2017pure, paesani2020near}. This has enabled demonstrations of quantum advantage with photons\cite{zhong2020quantum, zhong2021phase}, as well as the generation of large entangled states, with 12 photon Greenberger–Horne–Zeilinger entanglement\cite{zhong201812} and reconfigurable resource state generators of up to 8 qubits demonstrated\cite{adcock2019programmable, vigliar2021error}.
However, these demonstrations are limited by photon generation efficiency, with the mutliphoton data rate proportional to $R  p^n \eta^n$, for a $n$ photon system with  probability $p$ of photon generation, transmission $\eta$ and clockspeed $R$.
While $\eta$ can be improved by classical photonic engineering, nonlinear sources have an intrinsically limited $p$, as to avoid multiphoton emission, which scales as $O(p^2)$, and remains unchanged since the seminal quantum photonics experiments of the 1980s\cite{shih1988new, wu1986generation}.

Today, there are two promising approaches to achieving a near-deterministic single photon source.
The first is based on solid-state emitters, such as quantum dots \cite{somaschi2016near, tomm2021bright}, NV-center defects in diamond\cite{wan2020large}, or single molecules\cite{khasminskaya2016fully, toninelli2021single}. 
To reach $p=1$, these systems must be engineered to deterministically emit their single photons into a single guided mode.
Over the last two decades, the state of the art efficiency has risen to $p= 57$\% to optical fibre\cite{tomm2021bright} ($R=76$ MHz).
Emission of near-indistinguishable photons from array of solid-state emitters also remains challenging, as the wavelength and linewidth of the emitted photons are subject to variation based on the precise arrangement of emitter's atomic environment (for example, a quantum dot's shape and interaction with the surrounding crystal lattice).
Furthermore, while emission at standard telecommunications wavelengths has been demonstrated\cite{muller2018quantum}, telecom quantum dots are less mature, with purity and efficiency metrics lagging behind their counterparts in the near infra-red.

The second approach, known as photon source multiplexing, uses an array of probabilistic nonlinear photon pair sources\cite{pittman2002single, migdall2002tailoring}, and is the topic of this paper.
Here, the probability that at least one source among the array generates a photon pair, $p_h$, tends to unity as the number of sources, $m$, is increased.
Upon pair generation, one of the photons is detected to herald the presence of its partner, which can be dynamically switched to a single output mode.
In this way, if any source fires, a single photon is routed to the output.
Any loss experienced by the signal photon will reduce the photon output probability, therefore the switching network is required to have minimal loss, with unity probability reached with net transmission $\eta \rightarrow 1$ and $m\rightarrow \infty$.

Fully integrated photon source multiplexing is a clear path to scaling today's quantum photonic integrated circuits\cite{adcock2019programmable,vigliar2021error}, but remains a challenge due to the lack of low-loss high-speed switching technology on integrated quantum photonic platforms\cite{adcock2015advances}.
Furthermore, modern proposals for photonic quantum computing architectures\cite{bartolucci2021fusion, bartolucci2021creation, bourassa2021blueprint} utilise multiplexing extensively---conditional switching and measurement are crucial to both discrete and continuous variable proposed quantum computer architectures and for reaching loss- and error-tolerant thresholds\cite{gimeno2015three, bartolucci2021creation, bartolucci2021fusion}.

Today, chip- and fibre-based multiplexing systems have achieved enhancement factors in the $1$--$3$ range, with $m \leq 4$ sources utilised\cite{ma2011experimental, collins2013integrated, mendoza2016active, xiong2016active, francis2016all}, though enhancement factors of up to $28$ have been realised with up to $m=40$ time bins in bulk optics, benefiting from extremely low-loss Pockels cells\cite{kaneda2019high} based switches. 
Repetition rates remain in the $R=0.5$--$15$ MHz range\cite{ma2011experimental, kaneda2019high} across these demonstrations.
\ch{These relatively slow repetition rates are limited by available switching technology, while spontaneous photon generation has been demonstrated at clockspeeds up to 10 GHz\cite{zhang2017high}, offering a potential increase of three orders of magnitude. 
Indeed, for fully-integrated temporal multiplexing, GHz-speed switching is a requirement in order to reduce loss and footprint. 
Achieving these high clockspeeds may be possible with time- and wave-length division multiplexing of the pump\cite{zhang2015enhancing}, while further reductions can be achieved by multiplexing forward- and backward propagating optical modes of each nonlinear source\cite{xiong2013bidirectional}.}

In this paper, we present a temporally multiplexed single photon source based on a low-loss fibre switch and loop architecture.
Our resource efficient design multiplexes $m=11$ temporally independent (time bin) nonlinear sources, uses a single heralding detector, and achieves an enhancement heralded single photon emission of $4.5\pm0.5$.

\section{Multiplexed source}

Photon pair sources can be multiplexed across any set of orthogonal photonic modes.
For example, any combination of spatial\cite{collins2013integrated, mendoza2016active}, temporal\cite{kaneda2019high, mendoza2016active, xiong2016active}, or frequency\cite{zhang2015enhancing, joshi2018frequency} modes can be multiplexed, given the modes can be dynamically switched with high fidelity and low loss.
The loss of the mode conversion (switching network) determines the maximum possible enhancement and maximum number of effective sources, \ch{while the output probability is limited to the transmission of the output photon through the switching network to its application, $\eta_s$, known as the heralding efficiency.
Since any optical loss reduces output efficiency (as it also does with solid-state single photon sources), a truly deterministic source $p=1$ is infeasible.
However, loss-tolerant bounds for quantum computing and repeater protocols exists in the $>90$\% range\cite{bartolucci2021creation, bartolucci2021fusion, morley2018loss}.
}

Logarithmic tree structures provide the optimum enhancement for a given per-pass switching loss, with each photon traversing the (lossy) switch $\lceil \mathrm{log}_2(m) \rceil$ times\cite{bonneau2015effect}. \ch{These can be implemented in space with a physical logarithimic tree structure, or with $\mathrm{log}_2(m)$ delay lines with delays of length $2^n\tau$.}
However, these topologies require $O(m)$ switches and detectors, and thus inflate system size and cost dramatically, and feature increased propagation loss in routing.
Our temporal, loop-based architecture\cite{pittman2002single,  kaneda2019high}, instead only requires a single switch and detector, at the expense that the photons will instead traverse the switch $(m+1)/2$ times on average, and achieves similar performance for small numbers of bins.

In this work, we multiplex $m = 11$ temporally distinct sources within an output clock cycle, with photons generated by spontaneous four-wave mixing (SFWM) in a $1.2$ cm long silicon waveguide (Fig.~\ref{fig:fibre_plex_setup}). 
Here, bright pump pulses from a pulsed laser (1586 nm, 1 ps pulse duration, Calmar), are filtered with square-shaped WDM filters (0.44 nm, Opneti), before being injected into the waveguide via a grating coupler\cite{ding2014fully}.
Pairs of photons in the pump are then converted into pairs of photons at energy-conserving wavelengths symmetrically around the pump by the $\chi^{(3)}$ nonlinearity of the silicon waveguide.
The light is then extracted and the pump removed with further WDM filters at $1591.26$ and $1581.18$ nm, defining the herald and single photon output channels respectively, before detection in superconducting nanowire single photon detectors (SNSPDs, Quantum Opus), whose output is processed by a time-tagger (qutools quTAG) to recover photon rate and arrival time data.
\ch{The switch is an unpolarised, $1$ MHz rate, $2\times 2$ fibre optical switch (Photonwares NanoSpeed, proprietary), featuring an $8$ ns rise/fall time, sub-20 dB cross talk, and an $80$ ns short pulse capability. A high-voltage electrical driving system ($5$ V trigger) is also supplied.
}

To define a variable number of time-bins for multiplexing, pulses are picked from the $16$ MHz repetition rate laser using a variable optical attenuator (VOA).
Here, $m$ pulses, spaced by $\tau =125$ ns, are selected in a 2 $\upmu$s window, reducing the effective repetition rate to $8$ MHz.
This defines the output clock rate of the multiplexed source, $R = 500$ kHz, as well as the switch fibre loop length, which must delay the photons by $\tau$ per round trip ($25$ m fibre).
The switch and spliced fibre loop have a measured loss of $1$ dB per round trip, which combined with the source output loss determines a maximum photon probability enhancement of $4.5$, which is achieved in the limit of low base pair generation probability, i.e. average photon number $ \mu \approx p \approx 0.01$.

An FPGA state machine, which is clocked by the laser, controls both the multiplexing detect-switch-release logic, as well as generating the the pulse picking signal.
A fibre delay of $\delta = 200$ ns is included in the signal path before the fibre switch to compensate for the time taken to route the heralding photon to the detector and the resulting electronic signal back to the FPGA, which responds by activating the switch on the same clock cycle.

\begin{figure}[t!]
    \includegraphics[width=0.47\textwidth]{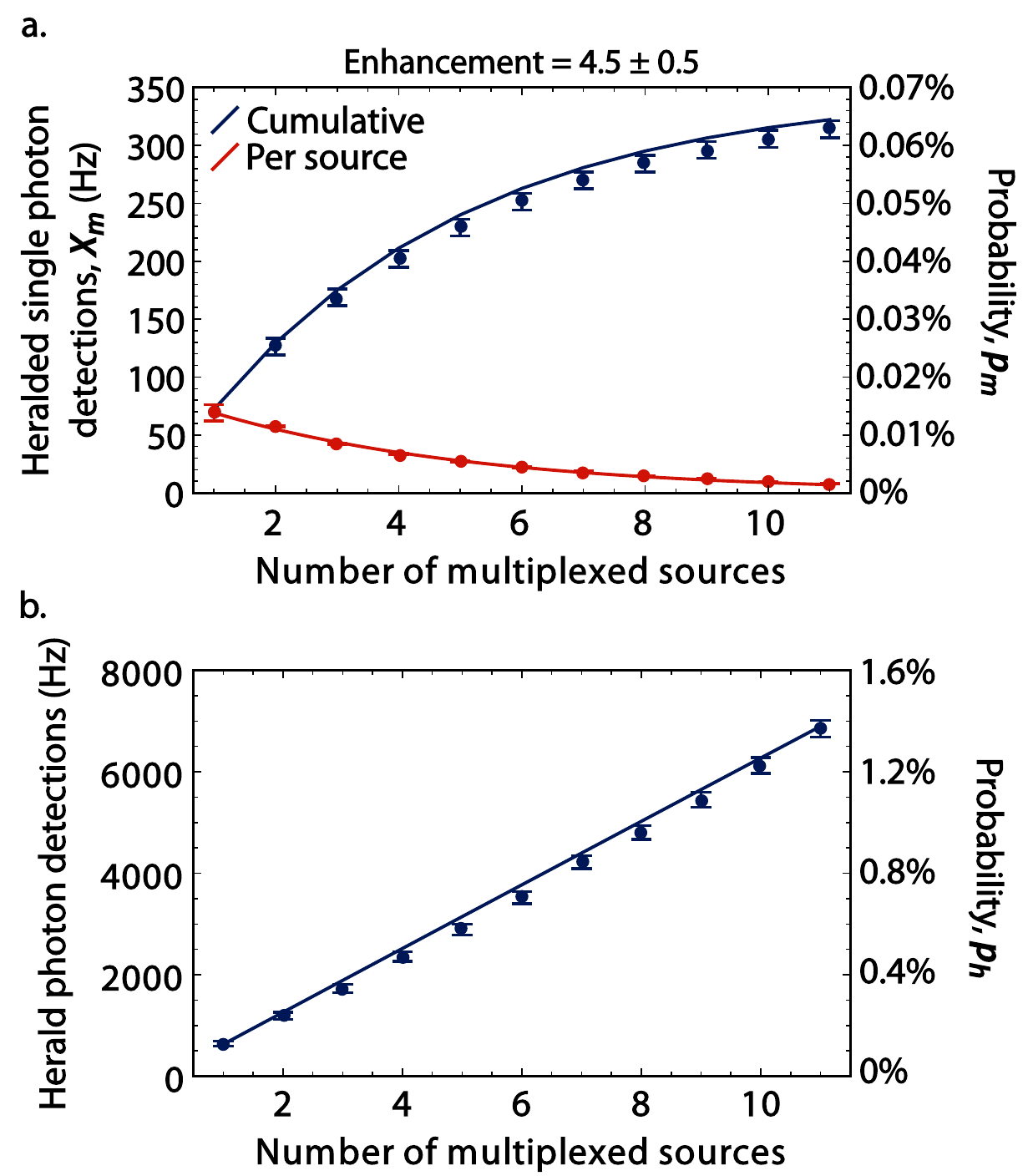}
    \centering
    \caption{Photon probability enhancement by temporal-loop multiplexing. \textbf{a.} Heralded photon probability and rate increase with the number of multiplexed sources (time bins). The contribution from each source decreases exponentially as photons in the $m^{\mathrm{th}}$ time bin must traverse the switch $m$ times, with $-1$ dB (79\%) transmission per trip. \textbf{b.} The probability and rate of detecting herald photons increases with number of multiplexed sources. Solid lines are a calibrated numerical model\cite{kaneda2019high}, using independently measured loss and brightness parameters.}
    \label{fig:results}
\end{figure}

\ch{While the switch's speed determines the output clockspeed, $R$, the pulse response time of the switch determines the minimum size for individual time bins, i.e. halving the pulse response time would allow twice as many time bins, $m$, to be utilised for the same $R$. 
Meanwhile, the loss in the switch determines the maximum attainable enhancement of the heralded single photon probability, with increased the number of bins resulting in diminishing returns toward a fixed limit corresponding to the net photon output transmission $\eta_s$ (Fig.~\ref{fig:results}). 
Our system is capable of multiplexing up to 16 sources without sacrificing overall output clockspeed, $R$, though to realise the potential of more than $m=11$ bins would require a switch with increased transmission (improvements beyond $m=11$ were insignificant).
In principle, the FPGA state machine supports up to 64 multiplexed sources, at the cost of reducing to $R$ to  $125$ kHz.
}

\subsection{Results}

We calibrate our base SFWM source to emit with a mean photon number of $\mu=CR/(S_hS_s)=0.009$ using the coincidence ($C$) to singles ($S$) ratio in the low power limit, where $g^{(2)}(0) \approx \mu$ and find the single photon transmissions $\eta_s = 0.113$, $\eta_h = 0.145$, including one pass through the switch and loop for the signal photon.
This low power is required to suppress multiphoton noise below $1$\%---required both for this calibration and for scalable quantum photonic protocols---and furthermore enables us to measure maximum attainable enhancement of the single photon probability. 

We measure the singles and coincidence rate for $m=\{1,\ldots,11\}$ bins and observe the contribution of each temporally independent photon pair source.
The contribution of each bin diminishes logarithmically, as photons generated in bin $k$ must traverse the switch $k$ times and the loop $k-1$ times.
Using $m=11$ multiplexed time bin sources, we achieve an enhancement, $E$, of $4.5 \pm 0.5$, of the heralded photon detection probability from $ p \cdot \eta_s \eta_h = 0.00014$ to $p_m = 0.00063 = E p \cdot \eta_s \eta_h$.
\ch{Here, $p \approx \mu = 0.009$ is the base generation probability in the source}.
We also match this with a model using our measured loss and brightness parameters for $m = \{1,\dots,11 \}$ time bins\cite{kaneda2019high} (Fig.~\ref{fig:results}).

For a mean photon number $\mu = 0.18$, as used in ref.~\citen{kaneda2019high}, our model predicts an output probability of $p_m = 1.4$\%, with an enhancement of $3.8$ at $m=11$ bins---though here the multiphoton emission rate is unacceptably high for most quantum applications.
Because we multiplex the first-generated photon pair, increasing $\mu$ results in higher probability that photon pairs are generated in earlier time bins, where the output photon experiences more loss, reducing the relative enhancement.
%

Due to imperfect matching between the photon storage loop and the laser repetition period, $\Delta_{\tau} = 1.7$ ns, the output photons, which have a coherence length $T_c \approx 5$ ps, are temporally distinguishable if they were generated in different time bins.
However, our simple and inexpensive design lends itself to quantum key distribution systems where photon indistinguishability is not required, but increasing photon probability boosts secret key rate.
Finally, our system comprises entirely of guided-mode optics, and is therefore stable and miniaturisable with appropriate packaging of the photonic integrated circuit.
Thanks to this, the system is also modular. For example, a high spectral purity fibre- or waveguide-based source\cite{paesani2020near, lugani2020spectrally}, or more efficient switch, could be swapped in to the system.

\subsection{Outlook}

\ch{Our device's output clock speed, $R$, is limited to $500$ kHz by its driving electronics (Photonwares).
However, with spontaneous photon sources supporting up to at least $R = 10$ GHz\cite{zhang2017high}, up to three orders of magnitude improvement in photon rate is possible.
Future work, using an integrated switch, such as low-loss thin-film lithium niobate (TFLN)\cite{he2019high} will enable on-chip photon routing at potentially at tens of GHz switching speeds with CMOS compatible voltages\cite{wang2018integrated}, enabling multiplexing  within the recovery time of commercial SNSPDs.
Here, increasing the clock speed $R$, reduces propagation losses, and decreases effective switch loss, resulting in higher attainable output efficiencies.
Furthermore, group delay can be precisely calculated and accounted for with fine-tuning achieved via temperature stabilisation, for indistinguishable photon generation.
Hybrid silicon/TFLN integrated photonics\cite{He19NPLNmod}, achieved via wafer bonding techniques, should enable photon sources and multiplexing switch networks to be fabricated on a single chip, enabling orders of magnitude improvements in efficiency and rate.}

\ch{The heralded photon probability is held back by our system's relatively low heralding efficiency, and moderate switch loss.}
On-chip propagation and grating coupler losses comprise the bulk of the loss (around $-6$ dB), with the remaining loss occurring in the filters ($-0.6$ dB), detectors ($-1$ dB), and fibre connections (around $-0.8$ dB), resulting in net transmissions (including detection) of $\eta_h = 0.145$, $\eta_s = 0.113$, including one pass through the switch for the signal photon.
\ch{This results in a base heralded photon output rate of $X_b = R p \eta_s \eta_h = 69 \pm 7$ Hz, which is increased to $X_m = 312 \pm 7$ Hz when $m=11$ time bins are multiplexed.}

\ch{
In our system, a zero-loss switch would yield $p_m = \eta_s'$, $X_m = 71.5$ kHz as $m\rightarrow \infty$ where $\eta_s' = 0.143$ is the transmission of the filtered photon without the switch. Here, any increase in $R$, limited by the switch, provides a factor increase in multiplexed output rate $X_m$.
Decreasing the chip coupling losses, for example by utilising modern, sub-dB grating couplers would enable this system to achieve heralding efficiencies of greater than $\eta_s' = 0.6$ to the output fibre, which gives an upper bound of multiplexed source output probability for a lossless switch, $p_m = \eta_s'$, $X_m = 300$ kHz.
}

\section{Conclusion}

Efficient single photon sources are a key challenge in today's proposed photonic quantum computer architectures\cite{bartolucci2021fusion, bourassa2021blueprint} and quantum networks\cite{azuma2015all, bartolucci2021creation}, which utilise feed-forward and dynamic quantum circuits as a core feature.
In this work we have boosted the heralded photon emission of a silicon waveguide photon pair source by dynamically detecting and routing photon pairs generated in up to $11$ multiplexed time bins.
Our system achieved a factor $4.5 \pm 0.5$ enhancement of the base probability rate, with negligible change to the $g^{(2)}(0)=0.01$.
\ch{This work demonstrates the increased efficiency of multiplexed quantum photonic hardware via a compact, guided-mode approach, on the path to developing future fully-integrated devices for large-scale quantum technology.}

\section{Acknowledgements}
The authors would like to thank Kjeld Dalgaard for designing the FPGA state machine which controlled the experiment. 
We acknowledge funding from Villum Fonden Young Investigator project QUANPIC (ref.~00025298) and Danish National Research Foundation Center of Excellence, SPOC (ref.~DNRF123).

\section{Contributions}
JCA conducted the experiment, analysed the data and wrote the manuscript, JCA, DB and YD designed the experiment, and YD managed the project.

\bibliography{references}

\end{document}